\documentclass[aps,prd,preprint,nofootinbib]{revtex4}
\usepackage[utf8]{inputenc}
\usepackage{amsmath,amssymb} 
\usepackage{graphicx} 
\usepackage{hyperref}

\newcommand{\be}{\begin{equation}}
\newcommand{\ee}{\end{equation}}
\newcommand{\bea}{\begin{eqnarray}}
\newcommand{\eea}{\end{eqnarray}}
\newcommand{\non}{\nonumber}

\begin{document}

\author{Christian Dioguardi}
\email{christian.dioguardi@alumni.unitn.it}
\affiliation{Dipartimento di Fisica, Universit\`{a} di Trento,\\Via Sommarive 14, I-38123 Povo (TN), Italy}
\author{Massimiliano Rinaldi}
\email{massimiliano.rinaldi@unitn.it}
\affiliation{Dipartimento di Fisica, Universit\`{a} di Trento,\\Via Sommarive 14, I-38123 Povo (TN), Italy}
\affiliation{Trento Institute for Fundamental Physics and Applications (TIFPA)-INFN,\\Via Sommarive 14, I-38123 Povo (TN), Italy}

\title{A note on the linear stability of black holes in quadratic gravity}

\begin{abstract}
Black holes in $f(R)$-gravity are known to be unstable, especially the rotating ones. In particular, an instability develops that looks like the classical black hole bomb mechanism: the linearized modified Einstein equations are characterized by an effective mass that acts like a massive scalar perturbation on the Kerr solution in General Relativity, which is known to yield instabilities.  In this note, we consider a special class of $f(R)$ gravity that has the property of being scale-invariant. As a prototype, we consider the simplest case $f(R)=R^2$ and show that, in opposition to the general case, static and stationary black holes are stable, at least at the linear level. Finally, the result is generalized to a wider class of $f(R)$ theories.
\end{abstract}

\maketitle

\section{Introduction}

The challenging questions left unanswered by General Relativity (GR), like the physical interpretation of singularities or the consistent quantization of the field equations call for exploring more general constructions. A well-known extension of GR is the so-called $f(R)$ gravity, where the Einstein-Hilbert Lagrangian ${\cal L}=\sqrt{g}R$ is replaced by a generic function of the Ricci scalar $R$, i.e.  ${\cal L}=\sqrt{g}f(R)$. This wide class of theories has been explored in many contexts, from black hole physics to cosmology, especially inflation and dark energy  (see \cite{De_Felice_2010, Sotiriou_2010} for general reviews). 

A very general feature of $f(R)$ gravity is that it can be mapped to a standard scalar-tensor theory of gravity by means of a conformal transformation \cite{Steinwachs}. The field equations in $f(R)$ usually have higher-than-second order derivatives and are difficult to solve. By changing frame, however, the extra degree of freedom is encoded in a dynamical scalar field with a potential that depends on the analytic form of $f(R)$ and the equations of motion are again of second order at most. One very relevant example is given by the so-called Starobinski inflationary model $f(R)=R+R^2/(6M^2)$, where $M$ is the inflationary mass scale. When mapped to the Einstein frame, the theory becomes equivalent, in the high energy limit, to the Higgs inflation model, where the standard model Higgs field is non-minimally coupled to gravity and acts as the inflaton field \cite{Bezrukov:2007ep,Giudice:2014toa}. 

Modified $f(R)$ gravity counts numerous analytic black hole solutions but their stability   is difficult to assess as the perturbation equations are in general of fourth-order. Nevertheless there are ways to overcome this difficulty, as suggested in \cite{Myung_2013}, where the stability of the Kerr metric was investigated in the case when $f(R)$ is an analytic function of $R$. The main result of this work is that rotating black holes are unstable due to the presence of a massive graviton that generates a black hole bomb. The effective mass of the graviton depends explicitly on the analytic form of $f(R)$, namely
\bea\label{effmass}
\mathcal{M}^2 = \frac{f'(R) -  R f''(R)}{3f''(R)}\,,
\eea
where the prime indicates the derivative of $f$ with respect to $R$ and the quantity is computed for $R= 0$, since the Kerr black hole is Ricci flat \cite{Myung_2013}. For analytic forms of $f$, that is
\bea\label{taylor}
f(R)=\alpha_0+\alpha_1R+\alpha_2R^2+{\cal O}(R^3)
\eea
the effective mass $\mathcal{M}$ is always non-zero and proportional to $\alpha_1/\alpha_2$. In particular, the presence of a linear term in the expansion is crucial to have a non-zero mass. However, for the class of models of the form $f(R)=R^n$ ($n>1$) the effective mass vanishes if $R=0$ for $n\neq 2$ and for any $R$ for $n=2$. Since this class contains the Kerr metric as exact solution, in this note we investigate whether the instability persists in this class of models.

Such a class of modified gravity has attracted some attention, especially  the case $R^2$ since it is manifestly scale-invariant. Static and rotating black holes, both asymptotically flat or (anti)de Sitter ((A)dS) where investigated in many papers, see e.g. \cite{article,article2,article3,article4,Duplessis:2015xva,Edery:2019txq,Sultana:2018fkw,Dent:2016efw}. The conformal factor $\Omega$ that maps the $f(R)$ theory from the Jordan frame to its scalar-tensor counterpart in the Einstein frame is proportional to $f'(R)$ \cite{De_Felice_2010,Sotiriou_2010}. Therefore, if $\alpha_1=0$ (that is, there is no linear term in the expansion \eqref{taylor}) and the solution has $R=0$ the mapping is not possible. In other words, these theories do not have a scalar-tensor interpretation. In particular, this holds for $f(R)=R^2$ but it might hold for all scale-invariant gravity models \cite{Rinaldi:2018qpu}.

The plan of the present work is the following: in Sec. \ref{sec2} we quickly review the linearization of the Einstein equations in the case $f(R)=R^2$. In Sec. \ref{sec3} we discuss the static black hole solutions (both flat and asymptotically (A)dS) and then we examine the rotating case in Sec. \ref{sec4}. Finally, we draw some conclusions and open issues in Sec. \ref{sec5}.

\section{Linearized equations in pure quadratic gravity}\label{sec2}

Before discussing the linearized equations, we first need to find the unpertubed ones  by varying the action with respect to the metric:
\begin{equation}\label{eq1}
    S = \frac{\alpha}{36} \int d^4 x \sqrt{-g}\hspace{0.08cm} R^2
\end{equation}
where $\alpha$ is a dimensionless coupling, owing to the scale-invariance of the model (in vacuum). The result is 
\begin{equation}\label{eq2}
    2 R R_{\mu\nu} - \frac{1}{2}g_{\mu\nu} R^2 + 2\Big(g_{\mu\nu}\Box - \nabla_\mu \nabla_\nu \Big) R = 0\,.
    \end{equation}
The linearization is achieved by replacing in (\ref{eq2}) the perturbed metric tensor
\begin{equation}\label{eq3}
    g_{\mu \nu} = \Bar g_{\mu\nu} + h_{\mu\nu} \,,
\end{equation}
where $\Bar g_{\mu\nu}$ represents a non-dynamical background solution and $h_{\mu\nu}$ the perturbation\footnote{From now on barred quantities denote background fields.}. The Ricci scalar and the Ricci tensor expanded to the lowest order read, respectively
\begin{align}
& R = \Bar R + \delta R + {\cal O}(\delta R^2) \label{eq4}\,,\\
& R_{\mu\nu} = \Bar R_{\mu\nu} + \delta R_{\mu\nu}  + {\cal O} (\delta R_{\mu\nu}^2)\,, \label{eq5}
\end{align}
where
\begin{align}
    &\delta R =  \Bar \Box h - \Bar \nabla^{\rho} \Bar \nabla^{\sigma} h_{\rho \sigma} - h_{\rho\sigma} \Bar R^{\rho \sigma}\,,  \\
    &\delta R_{\mu\nu} = \frac{1}{2}\Big(\Bar \Box h_{\mu\nu} + \Bar \nabla{_\mu}\Bar \nabla_{\nu} h - \Bar  \nabla_{\mu}  \Bar\nabla^{\rho} h_{\nu\rho} -   \Bar \nabla_{\nu} \Bar \nabla^{\rho} h_{\mu\rho}\Big)\,.
\end{align}
To proceed further, it is convenient to distinguish between the cases $\Bar R = 0$ and $\Bar R \neq 0$, that is between Ricci flat or asymptotically locally (A)dS background solutions, respectively. As mentioned in the Introduction,  the subspace of asymptotically flat solutions cannot be mapped to the Einstein frame. In fact,  the conformal transformation from Jordan to Einstein frame for a general $f(R)$ theory is given by \cite{De_Felice_2010,Sotiriou_2010}
\begin{equation}\label{conformal}
    \tilde{g}_{\mu\nu} = \Omega^2 g_{\mu\nu}\,,
\end{equation}
where $\Omega^2 = f'(R)$. When $f(R)=R^2$, $\Omega^2 = \alpha R /(9 \Tilde M^2)$, which vanishes when $R=0$ preventing the mapping\footnote{Here, $\Tilde M$ is an arbitrary mass scale necessary to keep the conformal factor dimensionless.}. Therefore, to determine the stability of this class of solutions we have no choice and we need to perturb the higher-order equations in the Jordan frame. However, in the Ricci flat case, this is not as difficult as it appears. Indeed, after inserting \eqref{eq3} and \eqref{eq5} in \eqref{eq2} and setting $\Bar R = 0$, one finds
\begin{equation}\label{linflatR2}
    (2 \Bar{R}_{\mu\nu} + 2 \Bar g_{\mu \nu} \Bar{\Box} - 2\Bar{\nabla}_\mu \Bar{\nabla}_\nu) \delta R = 0\,.
\end{equation}
By using the trace of \eqref{linflatR2}, one finally arrives to the simple equation
\begin{equation}\label{wave}
    \Bar \Box \delta R = 0\,.
\end{equation}
Thus, we see that for Ricci flat backgrounds we have no propagating tensor degrees of freedom, which means that the tensor perturbations are not dynamical at the linear level. So the problem of the stability of asymptotically flat black holes is reduced to that of a massless scalar field on a fixed background.

In the case $\Bar R \neq 0$ instead, we have non-trivial equations for both tensor and scalar perturbations. In fact, by inserting \eqref{eq3} and \eqref{eq5} in \eqref{eq2}, we find
\begin{equation}\label{tensor}
    \delta R_{\mu\nu} - h_{\mu \nu}\Lambda + \frac{1}{3 \mathcal{M}^2}\Big(\Bar g_{\mu\nu} \Lambda + \Bar g_{\mu \nu} \Bar \Box - \Bar \nabla_\mu \Bar \nabla_\mu \Big)  \delta R - \frac{1}{2}\Bar g_{\mu \nu} \delta R= 0\,, 
\end{equation}
where $\mathcal{M}$ is defined by \eqref{effmass} and $\Lambda$ is implicitly given by
\bea \label{a}
 \Bar R_{\mu\nu} &=& \frac14 \Bar g_{\mu\nu}\Bar R \equiv \Bar g_{\mu\nu} \Lambda \,,
\eea
By taking the trace of equation (\ref{tensor}) and using the  definitions of ${\cal M}$ and $\Lambda$ we get once again a massless scalar  equation 
\begin{equation}\label{box}
    \Bar \Box \delta R = 0\,.
\end{equation}
Thus, the stability of asymptotically-(A)dS black holes is determined by both scalar and tensor perturbations.

\section{Stability of static spherically symmetric black holes}\label{sec3}

 It is well-known that in $f(R)$ gravity there is no corresponding Birkhoff theorem \cite{Sotiriou_2010} so the static spherically symmetric solution is not unique. The most general spherically symmetric static solution takes the form
\begin{equation}
    ds^2 = -\alpha(r) dt^2 + \beta(r) dr^2 + r^2 d\Omega^2\,,
\end{equation}
where, for a given $\alpha(r)$, one can find a corresponding  function $\beta(r)$ by using solution generating techniques. For instance one can impose constant curvature \cite{Calza:2018ohl}. In this work we  focus on  the analytic solutions with $\alpha(r) = 1/\beta(r)$ which correspond to a line element similar to the Reissner-Nordstr\"om one \cite{article2}
\begin{equation}\label{Reissner}
ds^2 = -\Big(1-\frac{2M}{r} + \frac{K}{r^2}\Big) dt^2 + \Big(1-\frac{2M}{r} + \frac{K}{r^2}\Big)^{-1} dr^2 + r^2 d\Omega^2\,,
\end{equation}
where both $M$, $K$ must be regarded as mere integration constants. In particular, the parameter $K$ is not associated to an electromagnetic stress tensor since we are considering vacuum solutions\footnote{A similar behaviour is present in the case of cosmological solutions: by imposing the flat Robertson Walker metric one finds a scale factor that interpolates between a radiation-dominated Universe and a de Sitter space.}. These black holes have the interesting property that, despite their temperature is non-zero, when associated to the usual surface gravity, the Wald entropy is exactly zero \cite{article2}.
As mentioned above, in the Ricci flat case we only have the scalar perturbation $\delta R$, which can be written as
\begin{equation}
\delta R (t,r,\theta,\phi) = e^{-i \omega t} A_l(r) Y_{lm}(\theta,\phi)
\end{equation}
where $\omega$ is in general a complex number, and $l$,$m$ are the usual indices of the spherical harmonics. By defining the tortoise coordinates $dr_* = \alpha(r)^{-1} dr$ and a new radial function $u_l(r) =  \frac{\alpha(r)}{r} A_l(r)$ we can turn equation (\ref{wave}) into the Schr\"odinger-like equation for the radial part
\begin{equation} \label{schroedingereq}
  \Big(  -\partial^2_{r_*} + V(r)\Big)  u_l= \omega^2 u_l\,,
\end{equation}
where
\begin{equation}\label{potential}
    V(r) = \Big(1 -\frac{2M}{r} + \frac{K}{r^2}\Big)\Big(\frac{l(l+1)}{r^2} + \frac{2M}{r^3} - \frac{2K}{r^4}\Big)\,.
\end{equation}
In general, if $V(r)$ is positive, the operator on the left hand side of (\ref{schroedingereq}) is positive definite and self-adjoint\footnote{Self-adjointness is defined with respect to the usual inner product in $L_2$ Hilbert space, see e.g. \cite{wald}.}. This implies \cite{criteria} $\Im(\omega) \leq 0$ (the imaginary part of $\omega$)  and the scalar field modes remain bounded at any $t$.
The potential defined in \eqref{potential} is always positive definite in the causal region outside the black hole horizon whenever $M^2 > |K|$, a condition that we impose to avoid the appearance of naked singularities. We conclude that static spherically symmetric black holes with line element (\ref{Reissner}) are stable in $R^2$. In particular, by taking the limit $K\rightarrow 0$, we conclude that the Schwarzschild solution is also stable in $R^2$.

We now consider the case when $\Bar R \neq 0$. This choice corresponds to asymptotically-(A)dS solutions. Again the solution is not unique so we impose $\alpha(r) = 1/\beta(r)$ and find the line element \cite{article2}:
\begin{equation}
    ds^2 = -\Big(1-\frac{2M}{r}-\frac{\Lambda}{3}r^2\Big) dt^2 + \Big(1-\frac{2M}{r}-\frac{\Lambda}{3}r^2\Big)^{-1} dr^2 + r^2d\Omega^2
\end{equation}
where $M$, $\Lambda$ must be regarded, again, as integration constants.
By repeating the same procedure of the previous case we find a Schr\"odinger equation with potential
\begin{equation}
    V(r) = \Big(1-\frac{2M}{r} - \frac{\Lambda}{3}r^2\Big)\Big(\frac{l(l+1)}{r^2}+\frac{2M}{r^3}-\frac{2}{3}\Lambda\Big)\,,
\end{equation}
which is again always positive\footnote{There is one exception for the monopole perturbation $l=0$ in the case $\Lambda > 0$. For this choice of parameters the potential turns negative at $r_h<r_0<r_c$ between the black hole and cosmological horizon. However, this does not imply instability because the potential is still bounded from below and we can still have $\Im(\omega) \leq 0$. Actually, Schwarzschild-De Sitter solution was proven to be stable also at non-linear level against a massless scalar wave \cite{dafermos2007wave}, so the negativity of $V(r)$ in this case causes no trouble.} in the causal region between the black hole horizon and the cosmological horizon. We conclude that the Schwarzschild-(A)dS solution is stable against the linear scalar perturbation (\ref{box}). 

As we have discussed in the case of $\Bar R \neq 0$ we also have a tensor perturbation obeying (\ref{tensor}). To analyze this case,  we rely on the decomposition of the tensor $h_{\mu\nu}$ in terms of odd $h_{\mu\nu }^{\rm odd}$ and even $h_{\mu\nu }^{\rm even}$ perturbations based on their behavior under parity transformation \cite{PhysRev.108.1063},\cite{PhysRevLett.24.737}.
In the case of odd perturbations  \eqref{tensor} simplifies because the scalar $\delta R$ transforms as an even-parity perturbation and does not contribute to odd ones. Hence, the equation reduces to the same equation one finds for first order tensor perturbations in GR with a cosmological constant. After an involved calculation one arrives to the Regge-Wheeler potential
\cite{adspert}
\begin{equation}
       V_{RW}(r) = \Big(1-\frac{2M}{r} -\frac{\Lambda}{3}r^2\Big)\Big(\frac{l(l+1)}{r^2} - \frac{6M}{r^3} \Big)\,.
\end{equation}

Note that this form of the potential only holds for $l>1$. In fact,  the perturbations with $l=0,1$ are gauge modes so they are not relevant for the stability analysis \cite{rezzolla}. Thus, even if the potential has a zero at $r = \frac{6M}{l(l+1)}$, it is hidden behind the black hole horizon for $l>1$  so it does not spoil the positive definiteness of $V_{RW}$ outside the horizon. Hence, it is guaranteed that the frequency of the modes satisfies $\Im(\omega) \leq 0$.

The even modes are trickier. In fact equation (\ref{tensor}) describes the  tensor perturbation $h^{\rm even}_{\mu\nu}$ coupled to the scalar perturbation $\delta R$, hence it cannot be rewritten in the form of a Schr\"odinger-like equation. It is convenient in this case to perform a conformal transformation to the Einstein frame. There is no trouble in doing so since the conformal factor is well-defined for $R \neq 0$ backgrounds. Thus, by means of the conformal transformation \eqref{conformal}, it is straightforward to show that action \eqref{eq1} is equivalent to
\begin{equation}\label{action}
S = \int d^4x \sqrt{-\Tilde{g}}\Big(\frac{\tilde{M}}{2}(\Tilde{R} - 2\Lambda) - \frac{1}{2}\Tilde{g}^{\mu\nu}\partial_\mu \phi \hspace{0.08cm} \partial_\nu \phi\Big)\,,
\end{equation}
where it is understood that all quantities with a tilde are evaluated in the Einstein frame and differ from those in the Jordan frame. Here, the scalar field $\phi \propto \ln f'(R)$ appears explicitly as a dynamical term in the Lagrangian.
Since we are interested in the linearized equations, we expand \eqref{action} up to second order and find
\begin{equation}\label{secondorder}
    S^{(2)} = S_{EH}^{(2)} -\frac{1}{2} \int d^4 x \sqrt{-\Bar{g}}\hspace{0.08cm}\Bar{g}^{\mu\nu} \partial_\mu (\delta\phi) \partial_\nu (\delta\phi)\,,
\end{equation}
where $S_{EH}^{(2)}$ is the Einstein-Hilbert action expanded to the second order\footnote{Until the end of the section all quantities are evaluated in the Einstein frame, we omit the tilde to avoid cumbersome notation.} and $\delta \phi$ is the linearized scalar degree of freedom obtained by expanding $\phi = \Bar \phi + \delta \phi$.
From the definition of the conformal transformation (\ref{conformal}) one can see that the background quantities are the same up to a multiplicative factor. However, for first order perturbations the conformal transformation acts as a field redefinition. Hence the equations of motion will be written in terms of new fields in the Einstein frame related to old ones in the Jordan frame through (\ref{conformal}).
Variation of \eqref{secondorder} with respect to $\delta \phi$ and $h_{\mu\nu}$ (the linearized tensor perturbation) gives the set of decoupled linearized equations
\begin{equation}
    \delta {R}_{\mu\nu} = {\Lambda} {h}_{\mu\nu}\,,
\end{equation}
\begin{equation}
    \Bar \Box \delta\phi = 0\,,
\end{equation}
where $\delta R_{\mu \nu}$, $\Lambda$ and $h_{\mu\nu}$ are defined in Einstein frame and differ from those in the Jordan frame. We see that the problem of tensor perturbation is reduced once more to linearized GR with a cosmological constant. Thus, in the case of even perturbations the potential is the well-known Zerilli potential \cite{PhysRevLett.24.737}
\bea\non
    V_Z &=& \frac{2\alpha(r)}{r^4 T^2}\Big[r^2\Big(\frac{(l+2)(l-1)}{2}\Big)^2\Big(\frac{l(l+1)}{2}+\frac{2M}{r}\Big)+ \\\non \\
    &+& 9M^2 \Big(\frac{(l+2)(l-1)}{2} +\frac{M}{r} - \frac{\Lambda r^2}{3}  \Big)  \Big]\,,
\eea
where 
\begin{equation}
T = \frac{3M}{r} + \frac{(l+2)(l-1)}{2}\,.    
\end{equation}
This potential is always positive so we have $\Im(\omega)\leq 0$. Thus, we conclude that Schwarzschild-(A)dS solutions are stable against scalar and tensor perturbation in $R^2$ gravity.

\section{Stability of stationary black holes}\label{sec4}

As before, for stationary black holes, it is again  to discuss the two cases $\Bar R = 0$ and $\Bar R \neq 0$ separately. We begin with the former and we consider the Kerr metric, as it is an analytic solution of $R^2$ gravity \cite{article3}. The Kerr line element describes a rotating black hole in Boyer-Lindquist coordinates
\begin{equation}
    ds^2 = -\frac{\Delta}{\rho^2}(dt - a\sin^2\theta d\phi)^2 +\frac{\rho^2}{\Delta} dr^2 + \rho^2 d\theta^2 +\frac{\sin^2 \theta}{\rho^2}[a dt -(r^2+a^2)d\phi]^2\,,
\end{equation}
where $\Delta = r^2 - 2Mr +a^2$ and $\rho^2 = r^2 + a^2\cos^2 \theta$. Here $M$ is a length representing the mass parameter of the black hole and $a$ the angular momentum per unit mass.

For general $f(R)$ theories Kerr black holes have been proven to be unstable due to superradiance \cite{Myung_2013}. To show how the instability mechanism works we consider an impinging wave in the form
\begin{equation} \label{axisymmetric}
    \psi(t,r,\theta,\phi) = e^{-i\omega t + im\phi } A_{lm}(r) S_{lm}(\theta, \phi)\,,
\end{equation}
with complex frequency $\omega$, which scatters with a rotating object. If the condition of superradiance holds \cite{zeldovich}, that is the real part of $\omega$ satisfies
\begin{equation}\label{superradiance}
    \Re(\omega) < m\Omega\,,
\end{equation}
then the scattered wave absorbs energy from the rotating object and it is consequently amplified. In (\ref{superradiance}) $m$ is the multipole index of the impinging wave and $\Omega$ the angular velocity of the rotating object. If a mechanism for trapping the wave in the vicinity of the rotating object is provided then the superradiant modes yield an instability called black hole bomb. It was shown \cite{BHbomb_1} that in the case of Kerr black holes if a scalar wave $\Psi$ of the form \eqref{axisymmetric} with mass $\mu$ and satisfying the usual Klein-Gordon equation
 \begin{equation}
     (\Bar \Box - \mu^2)\Psi = 0 \,,
 \end{equation}
scatters on a rotating black hole, it can trigger the black hole bomb mechanism. The conditions under which this happens are the superradiance condition (\ref{superradiance}) together with \cite{Hod_2012}
\begin{equation}\label{condition}
    \frac{\mu^2}{2}<\omega^2<\mu^2\,.
\end{equation}
In fact, in this case the scalar wave is trapped and scatters back and forth between the black hole horizon and the potential barrier generated by the mass of the scalar field. This process extracts energy from the Kerr black hole until the potential barrier is destroyed. The second inequality in (\ref{condition}) follows from imposing bound state boundary conditions for the radial part of the scalar wave
\begin{equation}\label{bound}
\tilde{A}_{lm} \sim e^{\sqrt {\mu^2 -\omega^2 } r_*}, \hspace{1cm} r_*\rightarrow \infty
\end{equation}
where we have defined a new radial function $\Tilde{A}_{lm} = (r^2+a^2)^{1/2}A_{lm}$ and changed the radial coordinate to the tortoise coordinate $dr_* = (r^2 + a^2) \Delta^{-1} dr $ .

In the case of $R^2$ gravity, from \eqref{wave} we notice that the scalar perturbation $\delta R$ is massless so $\mu = 0$. Hence, from \eqref{bound} it follows that no bound state can be formed: the wave scatters with the black hole and then escapes to infinity. We can generalize this result to a larger class of $f(R)$ theories as follows. The definition of the effective mass of the scalar degree of freedom in $f(R)$ theories is given by (\ref{effmass}). 
Since we are considering the Kerr metric, we have $\Bar R = 0$ and (\ref{effmass}) reduces to
\begin{equation}\label{masseff}
    \mathcal{M}^2 = \frac{f'(0)}{3f''(0)}\,.
\end{equation}
Then (\ref{masseff}) vanishes in general for any analytic function $f(R)$ expanded as in (\ref{taylor}) with $\alpha_1 = 0$ and $\alpha_2 \neq 0$. This implies that the Kerr solution is not unstable: for any analytic $f(R)$ with $\alpha_1 = 0$ and $\alpha_2 \neq 0$ the black hole bomb mechanism cannot be triggered. This represents an exception with respect to the result found in \cite{Myung_2013} and can be considered the main result of this work.

Finally, we briefly consider the case $\Bar R \neq 0$. As in GR with a cosmological constant, the stationary solution is given by the class of Kerr-(A)dS black holes. In the case of Kerr Anti-de Sitter it has been shown \cite{Cardoso_2004, PhysRevD.80.084020} that in the limit of small black holes the boundary conditions imposed by the geometry are equivalent to a mirror placed at $r_0 \sim \ell$, where $\ell^2 = |\Lambda|/3$. Hence a trapping potential exists and the scalar wave $\delta R$ can generate an instability. The same might hold for Kerr-De Sitter solutions. The cosmological horizon effectively acts as a mirror and forces the scalar wave to scatter back and forth between the black hole horizon and the cosmological horizon. In this case, the superradiance condition equivalent to (\ref{superradiance}) is given by \cite{Zhang_2014}:
\begin{equation}
    m\Omega_c < \Re(\omega) < m\Omega_h 
\end{equation}
where $\Omega_h$ is the angular velocity of the black hole horizon and $\Omega_c$ is the cosmological horizon one. 
However, the only way to confirm the presence of this instability is to explicitly check the quasi-normal modes of the scalar field and look for those with $\Im(\omega)>0$, which is left to future work.

\section{Conclusions}\label{sec5}

In this note, we have re-examined the issue of the stability of black holes in $f(R)$ gravity. Generically, these are unstable at the linear level because there exist scalar perturbations that acquire a mass, which triggers the black hole bomb mechanism already known in standard GR. However, there are classes of scale-invariant modified gravity models, among which the simplest representative is $f(R)=R^2$, for which any effective mass that appears perturbatively would break scale-invariance\footnote{More sophisticate scale-invariant models require at least a scalar field, see e.g. \cite{Rinaldi:2015uvu}, or quadratic terms like $R_{\mu\nu}R^{\mu\nu}$.}.  Thus one might suspect that also these black holes are unstable. Here, we have proven instead that the scale-invariance protects the rotating asymptotically flat black hole from instabilities, at least at the linear level and outside the horizon. In fact,  eventual instabilities inside the black hole, like, for instance, the mass inflation near the inner Cauchy horizon, must be treated with other methods \cite{poisson}.

\end{document}